\documentclass{amsart}

\usepackage{amsmath,amssymb,amsxtra}

\theoremstyle{plain}
\newtheorem{theorem}{Theorem}
\newtheorem{corollary}{Corollary}
\newtheorem{lemma}{Lemma}
\newtheorem{proposition}{Proposition}

\theoremstyle{definition}
\newtheorem{definition}{Definition}

\newcommand{\B}{\mathbb}
\newcommand{\C}{\mathcal}

\newcommand{\ga}{\alpha}

\newcommand{\gd}{\delta}
\newcommand{\eps}{\varepsilon}

\newcommand{\gl}{\lambda}

\newcommand{\gs}{\sigma}

\newcommand{\gv}{\vartheta}

\newcommand{\gz}{\zeta}

\newcommand{\gG}{\Gamma}
\newcommand{\gL}{\Lambda}

\newcommand{\gQ}{\Theta}

\newcommand{\nn}{\nonumber}

\newcommand{\res}[3]{\underset{{{#1}={#2}}}{\textnormal{Res}}\left\{#3\right\}}
\newcommand{\fp}[3]{\underset{{{#1}={#2}}}{\textnormal{FP}}\left\{#3\right\}}

\begin{document}

\title{General moment theorems for non-distinct unrestricted partitions}

\author{Michael Coons}
\author{Klaus Kirsten}
\address{Simon Fraser University,
Department of Mathematics, 8888 University Drive, Burnaby, BC
V5A-1S6, Canada}
\email{mcoons@sfu.ca}
\address{Baylor University, Department of Mathematics,
        One Bear Place, Waco, TX 76798, USA}
\email{{K}laus\_{Kirsten}@baylor.edu}


\maketitle

\begin{abstract}A well-known result from Hardy and Ramanujan gives an
asymptotic expression for the number of possible ways to express an
integer as the sum of smaller integers. In this vein, we consider
the general partitioning problem of writing an integer $n$ as a
sum of summands from a given sequence $\gL$ of non-decreasing
integers. Under suitable assumptions on the sequence $\gL$, we
obtain results using associated zeta-functions and saddle-point
techniques. We also calculate higher moments of the sequence $\gL$
as well as the expected number of summands. Applications are made
to various sequences, including those of Barnes and Epstein types.
These results are of potential interest in statistical mechanics
in the context of Bose-Einstein condensation.
\end{abstract}

\section{Introduction}

The first significant ideas dealing with the theory of partitions
can be attributed to Euler \cite{Euler1} and were published in
1748. The next important milestone of partition theory was laid in
1918, when Hardy and Ramanujan \cite{HarRam1}, using quite
involved combinatorics, produced their celebrated theorem, which
gives the asymptotic result for the number of ways to express an
integer as the sum of lesser integers.

In the mid-1970's, the area of asymptotic analysis blossomed with
two of its primary works, those of Dingle \cite{Dingle} and Olver
\cite{Olv1}. Among the plethora of information in these two books
are some nice results concerning the method of steepest descent,
more commonly called the saddle-point method. Also at this time,
what is now one of the primary texts of partition theory,
Andrews', \emph{The Theory of Partitions} \cite{AndG1}, became
available. In 1975-76, Richmond used ideas of generating functions
\cite{lehm71p} and saddle-point methods to describe the moments of
certain types of partitions \cite{RichLB1,RichLB2}. In the first
of this two part series, \emph{The Moments of Partitions I}, using
asymptotic analysis instead of combinatorics, Richmond reproduced
the results of Hardy and Ramanujan \cite{HarRam1} as well as
calculating higher moments and variance. In his second paper,
\emph{The Moments of Partitions II}, Richmond went on to give
asymptotic results for the moments of a general sequence whose
associated zeta-function has only one singularity in the interval
$(0,1]$.

Though Richmond's results are remarkable, he does not consider the
situation where the sequence gives rise to a zeta-function with
arbitrarily many singularities at arbitrary values. Building upon
Richmond's results, and taking full advantage of the saddle-point
method as well as asymptotic analysis, the present paper addresses
that question.

As literature on zeta-functions is readily available, we give only
those definitions and propositions which are used. Background on
the Riemann zeta-function can be found in \cite{EdwHM} and
\cite{Titch}. Though the Barnes \cite{barn03-19-374,Barnes1} and
Epstein zeta-functions \cite{epst03-56-615,epst07-63-205} are less
well-known, they are of paramount importance in mathematical
physics; see, e.g., Actor \cite{acto94-230-303,acto95-43-141},
Dowker \cite{dowk94-162-633,dowk94-35-4989}, Elizalde
\cite{ElizE1,eliz95b,eliz94b} and Kirsten \cite{KirK1,KirK3}, for
a full reckoning. Details of a general zeta-function are given,
e.g., by Voros \cite{VorA1}.

After presenting the background, we derive what we call the
General Moment Theorems. The first of these theorems addresses the
question: given a sequence $\gL$ of non-decreasing integers, with
limited assumptions, how many ways are there to write a large
natural number $n$ as a sum of members from the sequence $\gL$?
The other General Moment Theorem addresses higher moments. We then
apply the General Moment Theorems to find the expected number of
summands.

We make applications to a variety of
sequences, starting with recreating the results of Richmond's first
paper, among other things re-deriving the famous Hardy and Ramanujan
theorem described above. The General Moment
Theorems are then applied to a sequence whose associated zeta-function has only one
singularity. In essence, this is what Richmond describes in his
second paper, except we assume no restriction on the location of the
singularity. We then reproduce the results of Nanda \cite{NanV1} for
two and three-dimensional Barnes-type partitions, and proceed to consider
two and three-dimensional Epstein-type partitions.

The advent of mathematical physics, specifically the area of
statistical mechanics in the context of Bose-Einstein
condensation, has brought partition theory to the forefront of
current research, see, e.g.,
\cite{holt98-270-198,holt01-300-433,nave97-79-1789}. It is because
of this context that we require the sequence $\gL$ to consist of
non-decreasing integers instead of increasing integers. More
specifically, we allow the sequence to contain the same integer a
finite number of times. In the physical context mentioned the
repetition of the same number represents {\it different} quantum
mechanical states with the {\it same} energy, so-called degenerate
states. Distributing a given amount of energy among the states
corresponds exactly to the partitioning problem considered here,
where the repeated numbers correspond to the different states with
the same energy. We hope that the results of this article will be
of interest to not only the physics community, but to number theorists as well.

\section{Zeta-functions}

\subsection{Basic zeta-functions}

In this section, we give the definitions and some properties of
specific zeta-functions needed for use in later sections.

\begin{definition} Let $s\in\B{C}$ with $\Re s>1$. The Riemann
zeta-function is defined as
$$\gz_\C{R}(s)=\sum_{n\in\B{N}}\frac{1}{n^s}.$$
\end{definition}
It is well known that a meromorphic extension of $\gz_\C{R} (s)$
to the whole complex plane can be constructed and that the only pole of $\gz_\C{R}
(s)$ is at $s=1$ \cite{EdwHM}.


%
%


\begin{definition} Let $s\in\B{C}$ with $\Re s>d$, and $c\in\B{R}$,
$\vec{r}\in\B{R}^d$ such that $c+\vec{m}\vec{r}> 0$ for all
$\vec{m}\in\B{N}_0^d$. The Barnes zeta-function is defined as
\cite{barn03-19-374,Barnes1}
$$\zeta_\C{B} (s,c|\vec{r})=\sum_{\vec{m}\in\B{N}_0^d}
\frac{1}{(c+\vec{m}\vec{r})^s}.$$ For $c=0$ it will be understood
that the summation only runs over $\vec m \in \B{N}_0^d-\{\vec 0\}$.
\end{definition}
For $\vec{r}=(1,1,\ldots,1,1):= \vec{1}$ we will use the notation
$\gz_\C{B}(s,c):= \gz_\C{B}(s,c|\vec{1})$, in which case we
have the following expansion \cite{barn03-19-374,Barnes1}:
\begin{proposition}\label{B degen} Let $\vec{r}=\vec{1}$ and $c>0$, then
$$\gz_\C{B}(s,c):= \gz_\C{B}(s,c|\vec{1})= \sum_{l=0}^\infty
e_l^{(d)}(c+l)^{-s},$$ where $$e_l^{(d)}=\left(\begin{matrix}
l+d-1\\ d-1\end{matrix}\right).$$
\end{proposition}
In the case $\vec r = \vec 1$, the above result indicates that the
Barnes zeta-function can be represented in terms of the so-called
Hurwitz zeta-function.
\begin{definition} Let $s\in\B{C}$ with $\Re s>1$ and $c\in\B{R}$
with $c>0$. The Hurwitz zeta-function is defined as
$$\zeta_H (s,c)=\sum_{n=0}^\infty \frac{1}{(n+c)^s}.$$
\end{definition}
The Hurwitz zeta-function can be meromorphically continued to the
whole complex plane and its only pole is located at $s=1$.

Barnes showed the following relation between the Hurwitz and
Barnes zeta-function \cite{barn03-19-374,Barnes1}:
\begin{corollary} The Barnes zeta-function $\gz_\C{B} (s,c)$ can be meromorphically
continued to the whole complex plane and for $c\in\B{R}$ with
$c>0$ one has
\begin{equation*}\label{Barnes-Hurwitz} \gz_\C{B}(s,c)=\sum_{k=1}^d
\frac{(-1)^{k+d}}{(k-1)!(d-k)!}B_{d-k}^{(d)}(c) \gz_H
(s+1-k,c).\end{equation*} Here $B_i^{(d)}(c):=
B_i^{(d)}(c|\vec{1})$ are the generalized Bernoulli polynomials \cite{norl22-43-121}
defined by
$$\frac{e^{-ct}}{\prod _{j=1}^d(1-e^{-r_jt})} =
\frac{(-1)^d}{\prod_{j=1}^d r_j} \sum_{n=0}^\infty
\frac{(-t)^{n-d}}{n!} B_n^{(d)} (c|\vec r).$$
\end{corollary}
Similar results can be derived for the case $c=0$. For example in
dimension $d=2$ one finds \begin{equation} \gz_{\C{B}_2}(s,0)=\zeta_{\C{R}}
(s-1) + \zeta_{\C{R}} (s),\label{twoBarn}\end{equation} whereas in dimension
$d=3$ the answer reads \begin{equation*} \gz_{\C{B}_3}(s,0)=\frac 1 2 \left(
\zeta_\C{R} (s-2) + 3\zeta_\C{R}(s-1) + 2 \zeta_\C{R}(s)\right).
\label{threeBarn}\end{equation*} Finally we will consider sequences related
to sums of squares of integers; the following zeta-function
will be useful.
\begin{definition} Define $Q(\vec{m},\vec{r})=r_1m_1^2+r_2m_2^2+\cdots+r_dm_d^2$.
Let $s\in\B{C}$ with $\Re s>\frac{d}{2}$, and $c\in\B{R}$,
$\vec{r}\in\B{R}^d$ such that $c+Q(\vec{m},\vec{r})> 0$ for all
$\vec{m}\in\B{N}_0^d$. The Epstein zeta-function \cite{epst03-56-615,epst07-63-205} is defined as
$$\zeta_\C{E} (s,c|\vec{r})=\sum_{\vec{m}\in\B{N}_0^d}
\frac{1}{(c+Q(\vec{m},\vec{r}))^s}.$$ In case $c=0$, it is
understood that the summation ranges over $\vec m \in
\B{N}_0^d-\{\vec 0\}$ only.
\end{definition}
For all cases meromorphic continuations of $\zeta_\C{E} (s,c|\vec
r)$ can be constructed; see, e.g.,
\cite{eliz94b,epst03-56-615,epst07-63-205,KirK1}. Because of the
complicated appearance, involving series over Bessel functions,
we do not display them explicitly.

%

\subsection{A general zeta-function}

In this section we first introduce the general type of sequences
we want to use, giving them certain basic restrictions, then we
examine residues and particular values of the zeta-function
associated with the sequence.

Throughout this paper we denote by $\gL$ a sequence that is a
nondecreasing sequence of natural numbers such that the
following hold.
\begin{enumerate}
\item[(i)] $1\in\gL$.
\item[(ii)] The partition function,  \begin{equation} \gQ
(t)=\sum_{\gl\in\gL} e^{-\gl t},\end{equation} converges for
$t>0$.
\item[(iii)] For $t\to 0^+$, $\gQ(t)$ admits  a full asymptotic expansion
\begin{equation}\label{Gen Asym}\gQ(t)\sim \sum_{n\in\B{N}_0}
A_{i_n} t^{i_n},\end{equation} where $i_n \in \B{R}$ with
$i_{n+1}> i_n$, $i_0<0$ and where $i_n\to\infty$ as $n\to\infty$.
Later on, we will occasionally also use the notation
$-i_n=\mu_n$.
\end{enumerate}
The fact that we restrict $\gL$ to be a sequence of natural numbers is due to the fact that we analyze partitioning problems
of natural numbers. In writing down the restrictions (ii) and (iii) we follow \cite{VorA1}, where it is shown that these requirements
lead to the well-defined spectral functions considered in the following. In particular, with these restrictions on the sequence $\gL$ of numbers, we
are now in a position to define a  general zeta-function (of
$\gL$-type) as the following.
\begin{definition} Let $\gL$ be a sequence as described above,
and let $s\in \B{C}$ with $\Re s> \mu_0$. We define the general
$\gL$-type zeta-function as \begin{equation*} \gz_\gL (s)=
\sum_{\gl\in\gL}\frac{1}{\gl^s}.\end{equation*}
\end{definition}
Typically, zeta functions are built from eigenvalues of an elliptic (pseudo) differential operator \cite{eliz95b,SeelRT3},
but it remains a perfectly viable spectral function in the given context \cite{VorA1}.

For our later considerations, we will only need to know residues
and particular values of $\gz_\gL$. To find these, we will use the
standard integral representation
\begin{equation*} \gz_\gL(s)=\frac{1}{\gG(s)}\int_0^\infty
t^{s-1}\gQ(t)dt.\end{equation*}
For the residues, and values of $\gz_\gL(s)$ at $s=-n$, only the
small-$t$ behavior of the integrand is relevant and we may focus
our examination on the function
\begin{equation*} L(s)=\frac{1}{\gG(s)}\int_0^1 t^{s-1}\gQ(t)dt,
\end{equation*}
realizing that the residues, and values of $\gz_\gL(s)$ at $s=-n$,
are precisely those of $L(s)$.

Formally substituting \eqref{Gen Asym} into the above equation and
performing the integration yields the expression
\begin{equation*}\label{Gen L} L(s) = \frac{1}{\gG(s)}
\int_0^1 t^{s-1}\sum_{n\in\B{N}_0} A_{i_n} t^{i_n}dt
=\frac{1}{\gG(s)}\sum_{n\in\B{N}_0} A_{i_n} \frac{1}{s+i_n}.
\end{equation*}
This formal calculation can be made precise and so allows to
conclude the following propositions \cite{SeelRT3,VorA1}:

\begin{proposition} For $i_n \neq k\in \B{N}_0$, the residues of $\gz_\gL (s)$ occur
at $s=-i_n$, and furthermore,
\begin{equation}\label{Gen Res} \res{s}{-i_n}{\gz_\gL(s)}=\frac{A_{i_n}}{\gG(-i_n)}.\end{equation}
\end{proposition}

\begin{proposition} For $n\in\B{N}_0$,
\begin{equation*} \gz_\gL(-n)=(-1)^n n! A_{n}.
\end{equation*}
\end{proposition}

\section{General Moment Theorems}

\subsection{General results for all $k$}

We start this section by defining moments of partitions.

\begin{definition} Let $p_\gL(n,m)$ be the number of partitions of $n$
into $m$ summands where each summand is a member of $\gL$. The
$k$-th moment of $p_\gL(n,m)$ is denoted by $t_\gL^k(n)$, and is
defined by
$$t_\gL^k(n)=\sum_{m\in\B{N}_0} m^kp_\gL(n,m),$$
where $p_\gL(n,m)=0$ for $m>n$ and $p_\gL(n,0)=0$.
\end{definition}

For small $n$, the above definition is sufficient for finding
values of $t_\gL^k(n)$ by hand or using a little computer
program. For large $n$, the calculation is much more difficult; we
must find a more reasonable way to compute $t_\gL^k(n)$. We
resolve this problem by first constructing a generating function,
then we continue by evaluating the coefficients of this generating
function; see \cite{lehm71p,RichLB1}.

Let us start by defining
\begin{equation} \label{Gen G(x,z)}
G_\gL(x,z)=\prod_{\gl\in\gL}(1-zx^\gl)^{-1}=\sum_{n\in\B{N}_0}\sum_{m\in\B{N}_0}p_\gL(n,m)
x^nz^m,\end{equation}
and
\begin{equation*} \gv=z\frac{\partial}{\partial z}.
\end{equation*}

Note that
\begin{equation*} \label{Gen gv^k G(x,z)}
\left.\gv^k G_\gL(x,z)\right|_{z=1}:= \gv^k
G_\gL(x)=\sum_{n\in\B{N}_0}\sum_{m\in\B{N}_0}m^kp_\gL(n,m)
x^n=\sum_{n\in\B{N}_0}t_\gL^k(n) x^n,\end{equation*} so that we
have constructed a generating function for $t_\gL^k(n)$.

Since $\gL$ contains only integers, only $n\in\B{N}_0$ occurs in
the summation. We may therefore apply Cauchy's formula for Laurent
series coefficients, so that for $\eps>0$ suitably chosen, we have
\begin{equation*}
t_\gL^k(n)=\frac{1}{2\pi i} \int_{C(0,\eps)} \gv^k
G_\gL(x)x^{-(n+1)}dx,\end{equation*}
where $C(0, \eps )$ is the circle of radius $\eps$ about $x=0$.
With the substitution $x=e^{-a}$, this easily becomes
\begin{equation}\label{gL sad int}
t_\gL^k(n)=\frac{1}{2\pi i} \int_{s.p.}e^{n(a+\frac{1}{n}\log \gv^k
G_\gL(e^{-a}))}da,\end{equation}
where $s.p.$ indicates a closed path that goes through the
saddle-point $a=\ga_k$ of the integrand.

The saddle-point $a=\alpha_k$ is found as a solution to the
equation \begin{equation} \frac d {da} \left( a + \frac 1 n \log \gv^k G_\gL
\left(e^{-a}\right) \right) =0. \label{speq1}\end{equation} As we will see
in the following, the large-$n$ expansion of the moments results
from a small-$|a|$ expansion of the saddle-point equation.

In order to evaluate (\ref{gL sad int}) as $n\to\infty$, that is,
as $|a|\to 0$, it will be necessary to find a more explicit form
of the saddle-point equation. We first simplify $\gv^k G_\gL
(e^{-a})$, making two cases, namely $k=1$ and $k\geq 2$. For
$k=1$,
\begin{align*} \gv
G_\gL(x,z)
&=z\frac{\partial}{\partial z}\left\{\prod_{\gl\in\gL} (1-zx^\gl)^{-1}\right\}\\
&=z\left(\sum_{\gl\in\gL}
\frac{x^\gl}{1-zx^\gl}\right)\left(\prod_{\gl\in\gL}
(1-zx^\gl)^{-1}\right)\\
\label{Gen Recur}&=G_\gL(x,z)\sum_{\gl\in\gL}
\frac{zx^\gl}{1-zx^\gl}.
\end{align*}
For convenience we denote
\begin{equation}
S_\gL(x,z)=\sum_{\gl\in\gL} \frac{zx^\gl}{1-zx^\gl} \quad
\mbox{and}\quad \left.\gv^k S_\gL (x,z) \right|_{z=1} := \gv^k S_\gL
(x).\label{defSgL}
\end{equation}
From the above string of equalities, we have $\gv G_\gL(e^{-a})=
G_\gL(e^{-a})S_\gL(e^{-a})$.

Repeating the above process for $k\geq 2$ yields \cite{RichLB1}
\begin{equation*} \gv^k
G_\gL(e^{-a})=G_\gL(e^{-a})S_\gL^{(k)}(e^{-a}),
\end{equation*}
with
\begin{multline}\label{Gen S^k} S_\gL^{(k)}(e^{-a})
=\sum \frac{k!}{b_1!b_2!\cdots b_k!}\left(\frac{\gv^0
S_\gL(e^{-a})}{1!}\right)^{b_1}\\ \times\left(\frac{\gv^1
S_\gL(e^{-a})}{2!}\right)^{b_2}\times
\cdots\times\left(\frac{\gv^{k-1}
S_\gL(e^{-a})}{k!}\right)^{b_k},\end{multline}
where the summation is over all solutions
$b_1,b_2,\ldots,b_k\in\B{N}_0$ of $b_1+2b_2+\cdots+kb_k=k$ (see \cite{JohnWP} for more details on sums like \eqref{Gen S^k}). For
readability, we denote the summation in (\ref{Gen S^k}) by $\sum
(\gv^0 S_\gL(e^{-\ga}),$ $\gv^1 S_\gL(e^{-\ga}),\ldots,\gv^{k-1}
S_\gL(e^{-\ga}))$. Using the above quantities, the saddle-point
equation (\ref{speq1}) reads \begin{equation} \frac d {da} \left( a + \frac 1
n \log G_\gL ( e^{-a}) + \frac 1 n \log S_\gL^{(k)}
( e^{-a}) \right) = 0 ,\nn\end{equation} or, more explicitly,
\begin{equation} n = \sum_{\lambda\in \gL} \frac \lambda {e^{a\lambda} -1} -
\frac { \frac  d {da} S_\gL^{(k)} \left( e^{-a} \right) }{S_\gL
^{(k)} \left( e^{-a} \right) } .\label{speq2}\end{equation} We next show
that for large $n$ the solution to this saddle-point equation is
unique.

At this point, we begin to make systematic use of the following
easily shown identity.

\begin{figure}[t]
\setlength{\unitlength}{0.14in} \centering
\begin{picture}(0,15)

\put(-15,7.5){\vector(1,0){30}} \put(0,0){\vector(0,1){15}}
\thicklines{\put(5,7.5){\line(0,-1){7}} \put(5,7.5){\vector(0,1){7}}
} \put(5.5,6.5){$\gs$} \put(15.5,7.25){$\Re t$} \put(-0.5,15.5){$\Im
t$} \put(0,7.5){\circle*{.25}}\put(-2,7.5){\circle*{.25}}
\put(-4,7.5){\circle*{.25}} \put(-6,7.5){\circle*{.25}}
\put(-8,7.5){\circle*{.25}} \put(-10,7.5){\circle*{.25}}
\put(-12,7.5){\circle*{.25}}
 \put(-3,6.5){$-1$} \put(-5,6.5){$-2$}
\put(-7,6.5){$-3$} \put(-9,6.5){$-4$} \put(-11,6.5){$-5$}
\put(-13,6.5){$-6$}

\end{picture}
\caption{The contour from $\gs-i\infty$ to $\gs+i\infty$ in the
complex plane.} \label{fig:contourgs}
\end{figure}
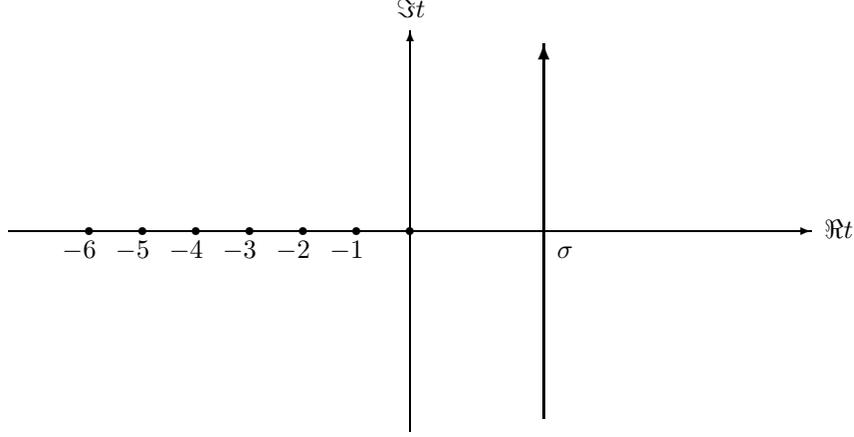

\begin{proposition}\label{(e^-z) int} Let $\gs>0$, $\gd>0$, and
$|\arg z|<\frac{\pi}{2}-\gd.$ Then
\begin{equation*}e^{-z}=\frac{1}{2\pi i}
\int_{\gs-i\infty}^{\gs+i\infty} z^{-t}\gG(t)dt ,\end{equation*}
where the limits of integration define the contour shown in Figure
\ref{fig:contourgs}.
\end{proposition}

Applying Proposition \ref{(e^-z) int} to the first term in
\eqref{speq2}, we have
\begin{align*} \sum_{\lambda \in \gL} \frac \lambda {e^{a\lambda} -1} =\sum_{\gl\in\gL}\sum_{l\in\B{N}} \frac{1}{2\pi i}
\int_{\gs-i\infty}^{\gs+i\infty} a^{-t}
\gl^{-(t-1)}l^{-t}\gG(t)dt,\end{align*}
which gives
\begin{equation*} \label{Gen n(ga)}  \sum_{\lambda \in \gL} \frac \lambda {e^{a\lambda} -1}
=\frac{1}{2\pi i}\int_{\gs-i\infty}^{\gs+i\infty}
a^{-t}\gz_\gL(t-1)\gz_\C{R}(t)\gG(t)dt,\end{equation*}
where the contour is similar to Figure \ref{fig:contourgs}, but
$\gs>\mu_0+1$ so that all of the residues of the integrand lie
to the left of $\gs$. This range for $\gs$ guarantees absolute
convergence of the resulting series and allows for an interchange
of summation and integration. In the right half-plane, the
integrand has simple poles at $t=\mu_0+1,\mu_1+1,\ldots, 1,0$, and
we find
\begin{equation} \label{Gen n(a)}
\sum_{\lambda \in \gL} \frac \lambda {e^{a\lambda} -1}
=\sum_{i=0}^d
\frac{\gz_\C{R}(\mu_i+1)}{a^{\mu_i+1}}\mu_iA_{-\mu_i}+\frac{A_0}{a}+
o\left( \frac 1 a\right) ,\end{equation}
where the $A_{k}$'s are defined by \eqref{Gen Asym}. The term $o
(1/a)$ summarizes subleading contributions as $a\to 0$, the
leading one of those behaving like $1/a^{1-\eps}$, $\eps >
0$, $\eps$ depending on the location of the right most pole of
$\zeta_\gL (s)$ on the negative real axis.

The analysis of the second term in (\ref{speq2}) follows along the
same lines. For $k=1$, using again Proposition \ref{(e^-z) int},
we write
\begin{equation} S_\gL(e^{-a}) =\frac{1}{2\pi i} \int_{\gs-i\infty}^{\gs+i\infty}
a^{-t}\gz_\gL(t)\gz_\C{R}(t)\gG(t)dt.\label{mb1}\end{equation} For
$k>1$ we need to focus on the terms $\gv^{s-1}S_\gL(e^{-a})$ for
$s>1$. Note that \eqref{defSgL} gives
\begin{equation} \label{Gen gv^s-1 S} \gv^{s-1}S_\gL(x)
=\sum_{\gl\in\gL}\frac{\sum_{j=1}^s
c_j^{(s)}x^{j\gl}}{(1-x^\gl)^s}=\sum_{j=1}^s
c_j^{(s)}\sum_{\gl\in\gL}\frac{x^{j\gl}}{(1-x^\gl)^s}=\sum_{j=1}^s
c_j^{(s)}S_\gL^{s,j}(x),\end{equation}
where
$S_\gL^{s,j}(x)=\sum_{\gl\in\gL}\frac{x^{j\gl}}{(1-x^\gl)^s},$ and
the $c_j^{(s)}$ are defined\footnote{Correcting a typo in
\cite{RichLB1}.} as in \cite{RichLB1} by $c_0^{(s)}=0,\
c_1^{(1)}=1,\ c_1^{(2)}=1,\ c_2^{(2)}=0,$ and for $s\geq 2$,
\begin{equation*} \label{cjs! coef} c_j^{(s+1)}=\begin{cases} jc_j^{(s)}+(s-j+1)c_{j-1}^{(s)},
& 1\leq j\leq s\\ 0, & j=s+1.\end{cases}\end{equation*} Also, note the
identity \begin{equation} \sum_{j=1}^s c_j^{(s)} = (s-1)!.\label{coefsum}\end{equation}

Using as before the substitution $x=e^{-a}$, we are interested in
sums of the form
\begin{equation*} S_\gL^{s,j}(e^{-a})=\sum_{\gl\in\gL}
\frac{e^{-a j\gl}}{(1-e^{-a \gl})^s}.\end{equation*}
A little arithmetic yields
\begin{equation*}
S_\gL^{s,j}(e^{-a})=\sum_{\gl\in\gL} \frac{e^{-a j \gl}}{(1-e^{-a
\gl})^s} =\sum_{\gl\in\gL} e^{-a j \gl}\sum_{l\in\B{N}_0}
e^{(s)}_l  e^{-a\gl l}=\sum_{\gl\in\gL}\sum_{l\in\B{N}_0}
e^{(s)}_l e^{-a\gl(l+j)},\end{equation*}
where $e^{(s)}_l=\left(\begin{matrix} l+s-1\\
s-1\end{matrix}\right)$. Applying Proposition \ref{(e^-z) int}
gives
\begin{align*} S_\gL^{s,j}(e^{-a})&= \sum_{\gl\in\gL}\sum_{l\in\B{N}_0}
 e^{(s)}_l  \frac{1}{2\pi i}
\int_{\gs-i\infty}^{\gs+i\infty} a^{-t}\gl^{-t}(l+j)^{-t} \gG(t)dt\\
&=  \frac{1}{2\pi i} \int_{\gs-i\infty}^{\gs+i\infty}
a^{-t}\left(\sum_{\gl\in\gL} \gl^{-t}\right)
\left(\sum_{l\in\B{N}_0}
 e^{(s)}_l (l+j)^{-t}\right) \gG(t)dt.\end{align*}
Since here we have to deal with Barnes
zeta-functions of different dimension, we adopt the notation that
$\gz_\C{B}^{(d)}$ is the Barnes zeta-function of dimension $d$.
With this new notation we find, recalling Proposition \ref{B
degen},
\begin{equation} S_\gL^{s,j}(e^{-a})=\frac{1}{2\pi i}
\int_{\gs-i\infty}^{\gs+i\infty} a^{-t}
\gz_\gL(t)\gz_\C{B}^{(s)}(t,j)\gG(t)dt.\label{mb2}\end{equation}
Although the small-$|a|$ expansion of (\ref{mb1}) and (\ref{mb2})
can be, and will be, obtained later on, at this stage let us
content ourselves with the following observation. The leading
$a\to 0$ behavior of $S_\gL ^{(k)} \left( e^{-a}\right)$, for
$l_k>0$ suitable, is seen to be of the form $a^{-l_k}$,
respectively $a^{-l_k} \log a$, depending on the location of
$\mu_0$. We therefore will have \begin{equation}\frac{ \frac d {da} S_\gL
^{(k)} \left( e^{-a} \right)} {S_\gL ^{(k)} \left( e^{-a} \right)
} = \frac {c_1 (k) + c_2 (k) / \log a} a ,\label{kdep1} \end{equation} the
numbers $c_1 (k)$ and $c_2 (k)$ depending on the location of $\mu
_0$. In all cases, it is seen that the $k$-dependent correction to
the saddle-point equation is of the order ${\C O} (1/a)$. This
will turn out to be of great importance for the analysis to
follow.

Using the small-$|a|$ expansion displayed in (\ref{Gen n(a)}) and
(\ref{kdep1}), the saddle-point equation (\ref{speq2}) allows us to
uniquely determine $a$ in terms of $n$, at least for large $n$.

With the saddle-point $\alpha_k$ known in terms of $n$, for large
$n$, we determine asymptotic answers for the moments by applying a
theorem of Olver (Thm 7.1, p.~ 126 of \cite{Olv1}) to \eqref{gL
sad int}. Noting that $$ \frac {d^2}{da^2} \left. \left( a + \frac
1 n \log \gv^k G_\gL \left( e^{-a} \right)
\right)\right|_{a=\alpha_k} = \left. - \frac 1 n \frac{dn} {da}
\right| _{a = \alpha _k},$$ this theorem takes the following form:
\begin{lemma} For $\ga_k$ the solution of \eqref{speq2}, as
$n\to\infty$, we have
\begin{equation}  \label{Gen t_(n) dn/da} t_\gL^k(n)=\frac{e^{n\ga_k}\gv^k
G_\gL (e^{-\ga_k})}{2\pi }\cdot
\left[\sqrt{\frac{-2\pi}{\frac{dn}{da}|_{a=\ga_k}}}+
\C{O}(n^{-3/2})\right].
\end{equation}
\end{lemma}
As indicated, $\alpha_k$ has to be thought of as being replaced by
its large-$n$ asymptotic expansion so that (\ref{Gen t_(n) dn/da})
represents a large-$n$ asymptotic expansion. This will be
explicitly done once we start looking at specific sequences
$\Lambda$ in Section 4.

The next important observation is that to leading order as
$n\to\infty$, the result for the moments is {\it independent} of
the $k$ used for the saddle-point $\alpha_k$.
\begin{proposition}\label{Gen recursion} For all $k$, as
$n\to\infty$, we have
\begin{equation*} t_\gL^k(n)= t_\gL^0(n)\cdot S_\gL^{(k)}(e^{-\ga_0}) [1+o (1) ].
\end{equation*}
\end{proposition}
\begin{proof}
Applying equation (\ref{Gen t_(n) dn/da}) amounts to evaluating
$e^{na+\log G_\gL (e^{-a})}$, $S_\gL ^{(k)}$ $(e^{-a})$, and
$dn/da$ at the saddle-points $a=\alpha_k$. We first note that the
saddle-point equation (\ref{speq2}), together with (\ref{Gen
n(a)}) and (\ref{kdep1}), imply \begin{equation*} \frac 1
{\alpha_k} = \left( \frac n { \zeta_\C{R} (\mu _0 +1) A_{-\mu _0}
\mu _0 } \right) ^{\frac 1 {\mu _0 +1}} [ 1 + o (1)] ,
\label{solsad1} \end{equation*} where $o(1)$ denotes terms that
vanish as $n\to \infty$. In particular, to leading order, $S_\gL
^{(k)} (e^{-\ga_k})$ and $dn/da|_{a=\ga_k}$ are independent of the
saddle-point used.

To show this independence for $e^{n\ga_k + \log G (e^{-\ga_k})}$
to the relevant order, considerably more work is necessary because
of the exponential magnifying factor. In order to show the
Proposition, we need to show that $n\alpha _k + \log G_\gL
(e^{-\alpha_k}) = n \alpha _0 + \log G_\gL ( e^{-\alpha _0 }) + o
(1)$ such that the difference due to the saddle-point chosen only
produces subleading order corrections. First, again from equations
(\ref{speq2}), (\ref{Gen n(a)}) and (\ref{kdep1}),
\begin{multline*} \frac 1 {\alpha_k}=\left( \frac n { \zeta_\C{R}
(\mu _0 +1) A_{-\mu _0} \mu _0 } \right) ^{\frac 1 {\mu _0 +1}}
\times\\
 \left( 1 - \frac 1 {\mu _0 +1} \,\,\frac
{\tilde c _1 (k) + c_2 (k) /\log n}{\zeta_\C{R} (\mu_0 +1) A_{-\mu _0}
\mu _0} \,\,\left( \frac n {\zeta_\C{R} (\mu_0 +1) A_{-\mu _0} \mu
_0}\right) ^{-\frac{\mu _0}{\mu _0 +1}} + \ldots
\right),\end{multline*} where the leading $k$-dependence of
the saddle-point solution has been depicted explicitly; $\tilde
c_1 (k)$ is determined from $c_1 (k)$ and $c_2(k)$. Using the
product representation of $G_\gL (e^{-a})$ from \eqref{Gen
G(x,z)}, we have
\begin{align}
\log G_\gL(e^{-a}) =\sum_{\gl\in\gL}\sum_{l\in\B{N}}\frac{e^{-a\gl
l}}{l}=\frac{1}{2\pi
i}\int_{\gs-i\infty}^{\gs+i\infty}a^{-t}\gz_\gL(t)\gz_\C{R}(t+1)\gG(t)dt.\label{parsum1}
\end{align}
Depicting only terms relevant for the leading $k$-dependence of
the contributions considered, we have $$\log G_\gL (e^{-\ga _k}) =
\ga_k ^{-\mu _0} \zeta_{\C R} (\mu_0 +1) A_{-\mu _0} + o (\ga _k
^{-\mu _0}). $$ Therefore, from equations (\ref{speq2}) and
(\ref{Gen n(a)}), we conclude
\begin{align*} n\alpha_k + \log G _\gL
(e^{-\alpha_k})  &= \tilde c _1 (k) + \frac 1 {\log n} c_2 (k) +
\ga_k^{-\mu_0} (\mu_0 +1) \zeta_\C{R} (\mu_0 +1) A_{-\mu _0}+ \ldots\\ &= 0
+ \ldots,\end{align*} that is, there is no $k$-dependence of this expression up
to the order $o (1)$, which shows the assertion.
\end{proof}
So in the following, saddle-point will always refer to $\ga :=
\ga _0$, which is a solution of \begin{equation} n = \sum_{\lambda\in\gL}
\frac \lambda {e^{\alpha \lambda } -1} .\label{speq3}\end{equation}

To exploit equations (\ref{Gen t_(n) dn/da}) and (\ref{Gen
recursion}) we first need a more complete expansion of equation
(\ref{parsum1}). The relevant integrand has simple poles at
$t=\mu_i$ for $i=0,1,\ldots,d$, and a double pole at $t=0$.
Calculation of the integral gives, as $\alpha \to 0$,
\begin{equation*}
\log G_\gL(e^{-\ga}) =
\sum_{i=0}^d\frac{\gz_\C{R}(\mu_i+1)}{\ga^{\mu_i}}A_{-\mu_i}-A_0\log\ga
+\gz_\gL'(0)+o(1).
\end{equation*}
Since $e^{n\ga}G_\gL(e^{-\ga})=e^{n\ga+\log G_\gL(e^{-\ga})}$, we
have the following lemma.

\begin{lemma} \label{Gen e^na G}For $\ga$ the solution of \eqref{speq3},
\begin{equation*}
e^{n\ga}G_\gL(e^{-\ga})=\ga^{-A_0}\exp\left[\sum_{i=0}^d
\frac{\mu_i+1}{\ga^{\mu_i}}
\gz_\C{R}(\mu_i+1)A_{-\mu_i}+A_0+\gz_\gL'(0)\right]
[1+o(1)].\end{equation*}
\end{lemma}

Changing focus to $\frac{dn}{da}|_{a=\ga}$, we note that this
quantity is not exponentiated in the solution of $t_\gL^k(n)$.
Thus, as mentioned, to obtain asymptotic results we need only
determine the leading order. With this in mind, from equation
\eqref{Gen n(a)}, as $\alpha \to 0$,
\begin{equation*}
\left.
-\frac{dn}{da}\right|_{a=\ga}=\frac{\mu_0(\mu_0+1)}{\ga^{\mu_0+2}}\gz_\C{R}(\mu_0+1)
A_{-\mu_0} \,\,\,[1+o(1)],\end{equation*}
so that
\begin{multline}\label{Gen 1/i[]}
\frac{1}{2\pi}\left[\sqrt{\frac{-2\pi}{\frac{dn}{da}|_{a=\ga}}}
+\C{O}(n^{-3/2})\right] =\left(2\pi\mu_0(\mu_0+1)\gz_\C{R}(\mu_0+1)
A_{-\mu_0}\right)^{-\frac{1}{2}}\\ \times\ga^{\frac{\mu_0}{2}+1}
[1+o(1)].
\end{multline}

With the main calculations behind us, we now begin to consider
specific $k$ values, making two cases; $k=0$ and
$k\geq 1$. The proceeding calculation brings us to our first general
moment theorem.

\subsection{General $\gL$-type: $k=0$}

To evaluate the 0-th moment of the general $\gL$-type,
$t_\gL^0(n)$, we apply \eqref{Gen t_(n) dn/da}, with $k=0$, along
with Lemma \ref{Gen e^na G} and \eqref{Gen 1/i[]}, to give the
following theorem.

\begin{theorem}\label{Gen Thm k=0} For $\gL$-type partitions,
\begin{multline*}
t_\gL^0(n)=  \left(2\pi\mu_0(\mu_0+1)\gz_\C{R}(\mu_0+1)
A_{-\mu_0}\right)^{-\frac{1}{2}}\ga^{\frac{\mu_0}{2}+1-A_0}\\
\times\exp\left[\sum_{i=0}^d \frac{\mu_i+1}{\ga^{\mu_i}}
\gz_\C{R}(\mu_i+1)A_{-\mu_i}+A_0+ \gz_\gL'(0)\right] [1+o(1)],
\end{multline*}
where $\ga$ is the solution of \eqref{Gen n(a)} and the $A_i$'s
are defined by \eqref{Gen Asym}.
\end{theorem}
Once $\alpha$ is replaced by its large-$n$ asymptotic expansion,
this determines the large-$n$ asymptotic expansion of $t_\gL^0
(n)$.

Let us recall that $t_\gL^0(n)$ gives the asymptotic result for the
number of partitions of an integer $n$ over the sequence $\gL$.

\subsection{General $\gL$-type: Case $k\geq 1$}\label{Gen Sec k>=1}

To evaluate the $k$-th moment of the general $\gL$-type for $k\geq
1$, $t_\gL^k(n)$, note that we need only evaluate
$S_\gL^{(k)}(e^{-\ga})$ and then apply Proposition \ref{Gen
recursion}.

It is evident from \eqref{Gen S^k} that within the calculation for
general $k\geq 1$, we will need the specific calculation for
$\gv^0S_\gL(e^{-\ga})=S_\gL(e^{-\ga})$.

Using Proposition \ref{(e^-z) int},
\begin{equation*} S_\gL(e^{-\ga}) =\frac{1}{2\pi i} \int_{\gs-i\infty}^{\gs+i\infty}
\ga^{-t}\gz_\gL(t)\gz_\C{R}(t)\gG(t)dt.\end{equation*}

To evaluate $S_\gL(e^{-\ga})$ we must distinguish between a few
cases. The above integral has different values depending on whether
$\mu_0<1$, $\mu_0=1$, or $\mu_0>1$. We treat these cases
independently.

For $\mu_0<1$, the leading pole is the simple pole of
$\gz_\C{R}(t)$ at $t=1$, and so
\begin{equation}\label{Gen k=1 mu<1} S_\gL(e^{-\ga}) =\frac{\gz_\gL(1)}{\ga}
\left[1+\C{O}\left(\ga^{1-\mu_0}\right)\right].\end{equation}

For $\mu_0=1$, the integrand has a double pole at $t=1$, and
\begin{align} S_\gL(e^{-\ga})
\label{Gen k=1 mu=1}=\frac{1}{\ga}\left(
\fp{t}{1}{\gz_\gL(t)}-A_{-1}\log\ga\right)
[1+\C{O}(\ga^{1-\mu_1})],\end{align} where $\fp{t}{1}{\gz_\gL(t)}$ denotes the finite part (or constant term) of the expansion of ${\gz_\gL(t)}$ around $t=1$.

For $\mu_0>1$, the integrand has a simple pole at $t=\mu_0$, and
\begin{align} S_\gL(e^{-\ga})
\label{Gen k=1 mu>1}=\frac{\gz_\C{R}(\mu_0)A_{-\mu_0}}{\ga^{\mu_0}}
\left[1+\C{O}\left(\ga^{\min(\mu_0-1,\mu_0-\mu_1)}\right)\right].\end{align}

Having completed the evaluation of $S_\gL(e^{-\ga})$, we now focus
on the terms $\gv^{s-1}S_\gL(e^{-\ga})$ for $s>1$. As shown
previously, see (\ref{Gen gv^s-1 S}) and (\ref{mb2}), the relevant
quantity to consider is
\begin{equation*} S_\gL^{s,j}(e^{-\ga})=\frac{1}{2\pi i}
\int_{\gs-i\infty}^{\gs+i\infty} \ga^{-t}
\gz_\gL(t)\gz_\C{B}^{(s)}(t,j)\gG(t)dt.\end{equation*}
Note that we need $k$ values for $s$, namely $s=1,2,\ldots,k$.
Previously in this section we considered $s=1$, now we turn our
attention to the following three cases for $s\geq 2$: $s<\mu_0$,
$s=\mu_0$, and $s>\mu_0$.

Let us remark, that although for each given case better error
terms could be given, in the generality considered there are many
cases necessary. Therefore, in order to make everything more
readable, we refrain from doing so.

For $s<\mu_0$, the leading term comes from the simple pole of
$\gz_\gL(t)$ at $t=\mu_0$, so that
\begin{align*} S_\gL^{s,j}(e^{-\ga})
&=\frac{\gz_\C{B}^{(s)}(\mu_0,j)A_{-\mu_0}}{\ga^{\mu_0}} \left[1+
o(1)\right].\end{align*}
Now we have, using the above and \eqref{Gen gv^s-1 S},
\begin{equation*}\label{Gen k>2
mu>s} \gv^{s-1} S_\gL(e^{-\ga}) =
 \ga^{-\mu_0}A_{-\mu_0}\sum_{j=1}^s c_j^{(s)}\gz_\C{B}^{(s)}(\mu_0,j)
\left[1+o(1)\right].
\end{equation*}

For $s=\mu_0$, the leading term comes from the double pole of
$\gz_\gL(t)\gz_\C{B}^{(s)}(t,j)$ at $t=\mu_0$. From
(\ref{Barnes-Hurwitz}) it is immediate that $$\res{t}{s}
{\gz_\C{B}^{(s)}(t,j)}= \frac 1 {\Gamma (s)},$$ and so
\begin{multline*} S_\gL^{s,j}(e^{-\ga})
=\frac{1}{\ga^s}\left(\fp{t}{s}{\gz_\gL(t)}+ A_{-s}
\fp{t}{s}{\gz_\C{B}^{(s)}(t,j)}\right.\\ \left.+\frac{A_{-s}}{\gG(s)}\left(
\psi(s)-\log\ga\right)\right)
\left[1+ o(1)\right].\end{multline*}
Thus, with (\ref{coefsum}),
\begin{multline*}\label{Gen k>2
mu=s} \gv^{s-1}
S_\gL(e^{-\ga})=\frac{1}{\ga^s}\left((s-1)!\cdot\fp{t}{s}{\gz_\gL(t)}+A_{-s}\left(
\psi(s)-\log\ga\right)+\right.\\
\left.+\sum_{j=1}^s
c_j^{(s)}\fp{t}{s}{\gz_\C{B}^{(s)}(t,j)}\right)\left[1+
o(1)\right].\end{multline*}

Finally, for $s>\mu_0$, the leading term comes from the simple pole
of $\gz_\C{B}^{(s)}(t,j)$ at $t=s$, so that
\begin{equation*} S_\gL^{s,j}(e^{-\ga})
=\ga^{-s}\gz_\gL(s) \left[1+o(1)\right],\end{equation*}
and so
\begin{equation}\label{Gen k>2 mu<s} \gv^{s-1}S_\gL(e^{-\ga}) =(s-1)!\cdot\ga^{-s}\gz_\gL(s)
\left[1+o(1)\right],\end{equation} where again (\ref{coefsum}) has
been used.

We are now in a position to evaluate $S_\gL^{(k)}(e^{-\ga})$ as
defined in \eqref{Gen S^k}; for the meaning of the notation $\Sigma$ followed by a $k$-component
expression consult the paragraph below \eqref{Gen S^k}. Again we must consider three cases:
$\mu_0<1$, $\mu_0=1$, and $\mu_0>1$.

If $\mu_0<1$, \eqref{Gen k=1 mu<1} and \eqref{Gen k>2 mu<s} give
\begin{equation*} \label{Gen S^k mu<1} S_\gL^{(k)}(e^{-\ga})=
\ga^{-k}\sum\left(\gz_\gL(1),\gz_\gL(2),2\gz_\gL(3),
\ldots,(k-1)!\gz_\gL(k)\right) [1+o(1)].\end{equation*}

If $\mu_0=1$, we have directly from \eqref{Gen k=1 mu=1} and
\eqref{Gen k>2 mu<s} that
\begin{multline*}\label{Gen S^k mu=1} S_\gL^{(k)}(e^{-\ga})=
\ga^{-k}\sum\left(\fp{t}{1}{\gz_\gL(t)}-A_{-1}\log\ga,\gz_\gL(2),2\gz_\gL(3),
\ldots,(k-1)!\gz_\gL(k)\right)\\
\times[1+o(1)].\end{multline*}

Finally, if $\mu_0>1$, then the leading term of
$S_\gL^{(k)}(e^{-\ga})$ comes from $b_1=k$. This implies
\begin{equation*} S_\gL^{(k)}(e^{-\ga})=
\left(S_\gL(e^{-\ga})\right)^k [1+o(1)],\end{equation*}
so that \eqref{Gen k=1 mu>1} gives
\begin{equation*} \label{Gen S^k mu>1} S_\gL^{(k)}(e^{-\ga})=
\ga^{-\mu_0k}A_{-\mu_0}^k\left[\gz_\C{R}(\mu_0)\right]^k [1+o(1)]
.\end{equation*}

We may now apply Proposition \ref{Gen recursion} to give the
following theorem.

\begin{theorem} \label{Gen Thm k>=1}For $k\geq 1$, and $\ga$ the solution of
\eqref{Gen n(a)}, the following results hold.
\begin{enumerate}
\item[(i)] If $\mu_0>1$, then
\begin{equation*} t_\gL^k(n)= t_\gL^0(n)\cdot
\ga^{-\mu_0k}A_{-\mu_0}^k\left[\gz_\C{R}(\mu_0)\right]^k [1+o(1)]
.\end{equation*}
\item[(ii)] If $\mu_0=1$, then
\begin{multline*} t_\gL^k(n)=t_\gL^0(n)\cdot
\ga^{-k}\sum\left(\fp{t}{1}{\gz_\gL(t)}-A_{-1}\log\ga,\right.\\
\left.\gz_\gL(2),2\gz_\gL(3),
\ldots,(k-1)!\gz_\gL(k)\right)[1+o(1)].\end{multline*}
\item[(iii)] If $\mu_0<1$, then
\begin{equation*} t_\gL^k(n)=t_\gL^0(n)\cdot
\ga^{-k}\sum\left(\gz_\gL(1),\gz_\gL(2),2\gz_\gL(3),
\ldots,(k-1)!\gz_\gL(k)\right) [1+o(1)].\end{equation*}
\end{enumerate}
\end{theorem}

We call Theorems \ref{Gen Thm k=0} and \ref{Gen Thm k>=1}, the
General Moment Theorems.

For $k=1$, the above theorem yields the following corollary.

\begin{corollary} \label{Gen Thm k=1} For $\gL$-type partitions, with $\ga$ the solution of \eqref{Gen n(a)}
and $t_\gL^0(n)$ as given in Theorem \ref{Gen Thm k=0}, we have:
\begin{enumerate}
\item[(i)] For $\mu_0<1$,
\begin{equation*} t_\gL^1(n)=t_\gL^0(n)\cdot
\frac{\gz_\gL(1)}{\ga}[1+o(1)] .
\end{equation*}
\item[(ii)] For $\mu_0=1$,
\begin{equation*} t_\gL^1(n)=t_\gL^0(n)\cdot\frac{1}{\ga}
\left( \fp{t}{1}{\gz_\gL(t)}-A_{-1}\log\ga\right) [1+o(1)].
\end{equation*}
\item[(iii)] For $\mu_0>1$,
\begin{equation*} t_\gL^1(n)=t_\gL^0(n)\cdot
\frac{\gz_\C{R}(\mu_0)A_{-\mu_0}}{\ga^{\mu_0}} [1+o(1)].
\end{equation*}
\end{enumerate}
\end{corollary}

\subsection{Expected number of summands}

In this section we give expressions for the expected number of
summands.

\begin{definition} The expected number of summands of a $\gL$-type
partition of an integer $n$, denoted by $m_\gL(n)$, is defined as
\begin{equation*} m_\gL(n)=\frac{t_\gL^1(n)}{t_\gL^0(n)}.\end{equation*}
\end{definition}

\begin{lemma}\label{m} The expected number of summands of a $\gL$-type
partition of an integer $n$ is
\begin{equation*} m_\gL(n)=S_\gL(e^{-\ga})
\left[1+o(1)\right],\end{equation*}
where $\ga$ is the solution of \eqref{Gen n(a)}.
\end{lemma}

\begin{proof} This result is an immediate consequence of Proposition
\ref{Gen recursion}.\end{proof}

In light of Lemma \ref{m}, a direct corollary of equations \eqref{Gen k=1 mu<1}, \eqref{Gen k=1
mu=1}, and \eqref{Gen k=1 mu>1}, is the following result.

\begin{theorem}\label{Gen m} For $\ga$ the solution of \eqref{Gen n(a)}, the
following assertions hold.
\begin{enumerate}
\item[(i)] If $\mu_0<1$, then
\begin{equation*} m_\gL(n)=\frac{\gz_\gL(1)}{\ga}
\left[1+o(1)\right].
\end{equation*}
\item[(ii)] If $\mu_0=1$, then
\begin{equation*}
m_\gL(n)=\frac{1}{\ga}\left(\fp{t}{1}{\gz_\gL(t)}-A_{-1}
\log\ga\right) \left[1+o(1)\right].
\end{equation*}
\item[(iii)] If $\mu_0>1$, then
\begin{equation*}
m_\gL(n)=\frac{\gz_\C{R}(\mu_0)A_{-\mu_0}}{\ga^{\mu_0}}
\left[1+o(1)\right].
\end{equation*}
\end{enumerate}
\end{theorem}

\section{Applications}\label{applications}

In this Section we apply the General Moment Theorems of the
previous section to a variety of special cases. We first reproduce
the results of Hardy and Ramanujan. Then we will examine the case of
one singularity at $\mu>0$. Furthermore, we examine
multidimensional applications in Barnes and Epstein type
sequences, where in each case we calculate higher moments and the
expected number of summands. The results frequently use the notation $\Sigma$ followed by a $k$-component expression. We
remind the reader that this notation is explained below \eqref{Gen S^k}.

\subsection{Hardy and Ramanujan}

Let $\gL = \B{N}$ such that $\zeta _\gL (s)= \zeta _\C{R} (s)$. This
is the case with one singularity at $\mu=1$ and we find the very
familiar results of Hardy and Ramanujan \cite{HarRam1} for the
asymptotic number of ways to express an integer $n$ as the sum of
lesser integers. The following theorems follow immediately from
the General Moment Theorems, see also \cite{RichLB1}.

\begin{lemma} For sufficiently large $n$,
\begin{equation*}\label{asym a}
\ga=\frac{\pi}{(6n)^{\frac{1}{2}}}- \frac 1 {4n}
+\C{O}(n^{-\frac{3}{2}}).
\end{equation*}
\end{lemma}

\begin{theorem}[Hardy and Ramanujan \cite{HarRam1}] The asymptotic number of ways
to partition $n$ over $\B{N}$ is
\begin{equation*}t_0(n)=
\frac{1}{4\sqrt{3} n} e^{\pi\sqrt{\frac{2n}{3}}}\left[1+\C{O}
(n^{-\frac{1}{2}})\right].\end{equation*}
\end{theorem}

\begin{corollary} Let $k=1$, then
\begin{equation*} t_1(n)=\frac{\sqrt{2}}{4\pi
\sqrt n}e^{\pi\sqrt{\frac{2n}{3}}}\left(\gamma+
\log\frac{\sqrt{6n}}{\pi}\right)\left[1+\C{O}(n^{-1/2})\right].
\end{equation*}
\end{corollary}

\begin{corollary} For $k\geq 2$,
\begin{multline*} t_k(n)=
\frac{1}{4\sqrt{3} n}
e^{\pi\sqrt{\frac{2n}{3}}}\left(\frac{\sqrt{6n}}{\pi}\right)^{k}
\times\\ \times\sum
\left(\gamma+\log\frac{\sqrt{6n}}{\pi},\gz_\C{R}(2),2 \gz_\C{R}
(3), \ldots,(k-1)! \gz_\C{R}(k)\right)\,\,\left[1+\C{O}
(n^{-\frac{1}{2}})\right].\end{multline*}
\end{corollary}

\begin{corollary} The expected number of summands of a Riemann
type partition of an integer $n$ is
\begin{equation*} m(n)=\frac{\sqrt{6n}}{\pi}\left(\gamma +\log
\frac{\sqrt{6n}}{\pi}\right)\left[1+\C{O}\left(n^{-\frac{1}{2}}\right)\right].\end{equation*}
\end{corollary}

%
\subsection{Zeta-functions with one singularity}

Let us suppose that we have a nondecreasing sequence of natural numbers
$\gL$, whose corresponding partition function $\gQ(t)$ admits the
full asymptotic expansion
\begin{equation*} \gQ(t)=\sum_{i=0}^\infty
A_{k_i}t^{k_i},\end{equation*}
in which $k_0<0$, and $k_i\geq 0$ for all $i>0$. Define
$\mu=-k_0$. Then the zeta-function associated with the sequence
$\gL$ has only one singularity on the positive real axis, namely
at $t=\mu$. Having met the assumptions of the General Moment
Theorems, we proceed to apply them to the above sequence $\gL$.

Note that within the previous section we could not easily solve
for $n$ in terms of the saddle-point $\ga$. Since we now have only
one singularity, the leading orders can be determined easily and
\eqref{Gen n(a)} gives
\begin{equation*}
\ga = \left(\frac{\gz_\C{R}(\mu+1)\mu
A_{-\mu}}{n}\right)^{\frac{1}{\mu+1}}+ \frac {A_0} {\mu+1}
\,\,\frac 1 n + o (1/n)
\end{equation*}

The General Moment Theorems now yield the following corollaries.

\begin{corollary} For sequences $\gL$ as described,
we have
\begin{multline*} t_\gL^0(n)=
\left(2\pi(\mu+1)\right)^{-\frac{1}{2}}\left(\mu\gz_\C{R}(\mu+1)A_{-\mu}
\right)^{\frac{1-2A_0}{2(\mu+1)}}
n^{\frac{2A_0-2-\mu}{2(\mu+1)}}\times\\
\times\exp\left[
\left(\frac{n}{\mu}\right)^{\frac{\mu}{\mu+1}}(\mu+1)\left(\gz_\C{R}(\mu+1)
A_{-\mu}\right)^{\frac{1}{\mu+1}}+\gz_\gL'(0)\right]
\,\,\,\left[1+o(1)\right].
\end{multline*}
\end{corollary}

For the higher moments, according to Theorem \ref{Gen Thm k>=1},
we need to distinguish three cases.
\begin{corollary} For sequences $\gL$ as described,
we have for $k\geq 1$ the following.
\begin{enumerate}
\item[(i)] For $\mu>1$,

\begin{equation*} t_\gL^k(n)= t_\gL^0(n)\cdot \left(\frac{n}
{\mu\gz_\C{R}(\mu+1)}\right)^{\frac{\mu k}{\mu+1}}
A_{-\mu}^{\frac{k}{\mu+1}}\left[\gz_\C{R}(\mu)\right]^k [1+o(1)].
\end{equation*}
\item[(ii)] For $\mu=1$,
\begin{multline*} t_\gL^k(n)= t_\gL^0(n)\cdot n^{\frac{k}{2}}\left(
\frac{\pi^2 A_{-1}}{6}\right)^{\frac{-k}{2}}\\
\times\sum\left(\fp{t}{1}{\gz_\gL(t)}-\frac{A_{-1}}{2}\log\left(\frac{\pi^2
A_{-1}}{6n}\right),\gz_\gL(2),2\gz_\gL(3),
\ldots,(k-1)!\gz_\gL(k)\right)\\
\times[1+o(1)]
\end{multline*}
\item[(iii)] For $\mu<1$,
\begin{multline*} t_\gL^k(n)= t_\gL^0(n)\cdot n^{\frac{k}{\mu+1}}\left(
\gz_\C{R}(\mu+1)\mu A_{-\mu}\right)^{\frac{-k}{\mu+1}}\\
\times\sum\left(\gz_\gL(1),\gz_\gL(2),2\gz_\gL(3),
\ldots,(k-1)!\gz_\gL(k)\right) [1+o(1)].
\end{multline*}
\end{enumerate}
\end{corollary}

Applying Theorem \ref{Gen m} gives the following corollary.

\begin{corollary}For sequences $\gL$ as described,
the following hold.
\begin{enumerate}
\item[(i)] If $\mu<1$, then
\begin{equation*} m_\gL(n)=\gz_\gL(1)\left(\frac{n}{\gz_\C{R}(\mu+1)\mu
A_{-\mu}}\right)^{\frac{1}{\mu+1}} \left[1+o(1)\right].
\end{equation*}
\item[(ii)] If $\mu=1$, then
\begin{equation*}
m_\gL(n)=\left(\frac{6n}{\pi^2 A_{-1}}\right)^{\frac{1}{2}}
\left(\fp{t}{1}{\gz_\gL(t)}-\frac{A_{-1}}{2} \log\frac{\pi^2
A_{-1}}{6n}\right) \left[1+o(1)\right].
\end{equation*}
\item[(iii)] If $\mu>1$, then
\begin{equation*}
m_\gL(n)=\gz_\C{R}(\mu )A_{-\mu}^{\frac 1 {\mu+1}}
\left(\frac{n}{\gz_\C{R}(\mu +1)\mu }\right)^{\frac{\mu}{\mu+1}}
\left[1+o(1)\right].
\end{equation*}
\end{enumerate}
\end{corollary}

%
%
%
%
%
%
%
\subsection{Barnes type moments}

We will now consider two special cases of Barnes type moments, the
two-dimensional and three-dimensional cases, with
$\vec{r}=\vec{1}$ and $c=0$. These results correspond to two and
three-dimensional oscillator assemblies considered by Nanda
\cite{NanV1}.

\subsubsection{Two-dimensional}

For the two-dimensional case, \eqref{Gen n(a)} gives
\begin{equation*} \ga=
\left(\frac{2\gz_\C{R}(3)}{n}\right)^{\frac{1}{3}}+\left(\frac{\gz_\C{R}(2)
}{3(2\gz_\C{R}(3))^{\frac{1}{3}} }\right)\frac 1
{n^{\frac{2}{3}}}-\frac{7}{36n}+\C{O}\left(n^{-\frac{4}{3}}\right).\end{equation*}
We use (\ref{twoBarn}) to evaluate relevant residues and
\begin{equation*}
\gz_{\C{B}_2}'(0,0)=-\frac{1}{2}\log 2\pi+\gz_{\C{R}}'(-1).
\end{equation*}

We may now apply the General Moment Theorems to produce the following
corollaries.

\begin{corollary}\label{Barnes 2} For the two-dimensional Barnes type sequence with $c=0$
and $\vec{r}=\vec{1}$,
\begin{multline*} t_{\C{B}_2}^0(n) =
\frac{(6\gz_\C{R}(3))^{-\frac{1}{2}}}{2\pi}\left(\frac{2\gz_\C{R}(3)}{n}\right)^{\frac{31}{36}}
\\
\times\exp\left[\frac{3(\gz_\C{R}(3))^{\frac{1}{3}}}{2^{\frac{2}{3}}}n^{\frac{2}{3}}
+\frac{\gz_\C{R}(2)}{2^{\frac{1}{3}}(\gz_\C{R}(3))^{\frac{1}{3}}}n^{\frac{1}{3}}-
\frac{(\gz_\C{R}(2))^2}{12\gz_\C{R}(3)}+\gz_\C{R}'(-1)\right]\\
\times\left[1+\C{O}\left(n^{-\frac{1}{3}}\right)\right].
\end{multline*}
\end{corollary}

Making the substitution $n'=n/\left(2\gz_\C{R}(3)\right)$, for
ease of comparison, the above corollary corresponds to that of Nanda \cite[Eq.~(34)]{NanV1}.

\begin{corollary}\label{Barnes 2k} For the two-dimensional Barnes type sequence with $c=0$,
$\vec{r}=\vec{1}$, and $k\geq 1$,
\begin{multline*} t_{\C{B}_2}^k(n) =
\frac{(6\gz_\C{R}(3))^{-\frac{1}{2}}}{2\pi}\left( \gz_\C{R} (2)
\right)^k \left(\frac{2\gz_\C{R}(3)}{n}\right)^{\frac{31-24k}{36}}
\\
\times\exp\left[\frac{3(\gz_\C{R}(3))^{\frac{1}{3}}}{2^{\frac{2}{3}}}n^{\frac{2}{3}}
+\frac{\gz_\C{R}(2)}{2^{\frac{1}{3}}(\gz_\C{R}(3))^{\frac{1}{3}}}n^{\frac{1}{3}}-\frac{(\gz_\C{R}(2))^2}{12\gz_\C{R}(3)}+\gz_\C{R}'(-1)\right]\\
\times\left[1+\C{O}\left(n^{-\frac{1}{3}}\right)\right].
\end{multline*}
\end{corollary}

Note that the case $k=0$ of Corollary \ref{Barnes 2k} is precisely
$t_{\C{B}_2}^0(n)$ of Corollary \ref{Barnes 2}. In general this is
true when the largest singularity is greater than 1; that is, in
this paper, $\mu_0>1$. Thus when considering the three-dimensional
Barnes case, we will give only the general result for
$k\in\B{N}_0$. But first we give the expected number of summands
as a consequence of Theorem \ref{Gen m}.

\begin{corollary} For the two-dimensional Barnes type partitions with $c=0$
and $\vec{r}=\vec{1}$,
\begin{equation*}
m_{\C{B}_2}(n)=\gz_\C{R}(2)\left(\frac{n}{2\gz_\C{R}(3)}\right)^{\frac{2}{3}}
\left[1+\C{O}\left(n^{-\frac{1}{3}}\right)\right].\end{equation*}
\end{corollary}

%

\subsubsection{Three-dimensional}

For the three-dimensional case, we evaluate \eqref{Gen n(a)} to
give
\begin{multline*} \ga= (3\gz_\C{R}(4))^\frac{1}{4} n^{-\frac{1}{4}}+
\left(\frac{3^{\frac{1}{2}}\gz_\C{R}(3)}{4(\gz_\C{R}(4))^{\frac{1}{2}}}\right)n^{-\frac{1}{2}}
\\+\left(\frac{8\gz_\C{R}(2)\gz_\C{R}(4)-3(\gz_\C{R}(3))^2}{3^{\frac{1}{4}}32(\gz_\C{R}(4))^{\frac{5}{4}}}\right)
n^{-\frac{3}{4}}-\frac{5}{32n}+\C{O}\left(n^{-\frac{5}{4}}\right).\end{multline*}

Using (\ref{threeBarn}) we evaluate relevant residues and
\begin{equation*}
\gz_{\C{B}_3}'(0,0)=-\frac{1}{2}\log
2\pi+\frac{3}{2}\gz_\C{R}'(-1)+\frac{1}{2}\gz_\C{R}'(-2).
\end{equation*}

We apply the General Moment Theorems to derive the following
corollaries.

\begin{corollary}\label{Barnes 3k} For the three-dimensional Barnes type sequence
with $c=0$, $\vec{r}=\vec{1}$, and $k\in\B{N}_0$,
\begin{multline*} t_{\C{B}_ 3}^k(n)
=\frac{(3\gz_\C{R}(4))^{-\frac{1}{2}}}{4\pi}\left(\gz_\C{R}(3)\right)^k
\left(\frac{3\gz_\C{R}(4)}{n}\right)^{\frac{25-24k}{32}}\\
\times\exp\left[4\gz_\C{R}(4)\left(\frac{n}{3\gz_\C{R}(4)}\right)^{\frac{3}{4}}+
\frac{3\gz_\C{R}(3)}{2}\left(\frac{n}{3\gz_\C{R}(4)}\right)^{\frac{1}{2}}\right.\\
\left.+\left(\gz_\C{R}(2)-\frac{3(\gz_\C{R}(3))^2}{8\gz_\C{R}(4)}\right)\left(\frac{n}{3\gz_\C{R}(4)}\right)^{\frac{1}{4}}
+C\right]\left[1+\C{O}\left(n^{-\frac{1}{4}}\right)\right],
\end{multline*}
where
$C=\frac{(\gz_\C{R}(3))^3}{8(\gz_\C{R}(4))^2}-\frac{\gz_\C{R}(2)\gz_\C{R}(3)}{4\gz_\C{R}(4)}
+\frac{3}{2}\gz_\C{R}'(-1)+\frac{1}{2}\gz_\C{R}'(-2)$.
\end{corollary}

Setting $k=0$, and making the substitution
$n''=n/\left(3\gz_\C{R}(4)\right)$, the above corollary corresponds
to that of Nanda \cite[Eq.~(51)]{NanV1}.

Now, Theorem \ref{Gen m} yields the following corollary.

\begin{corollary} For the three-dimensional Barnes type partitions with $c=0$
and $\vec{r}=\vec{1}$,
\begin{equation*}
m_{\C{B}_3}(n)=\gz_\C{R}(3)\left(\frac{n}{3\gz_\C{R}(4)}\right)^{\frac{3}{4}}
\left[1+\C{O}\left(n^{\frac{-1}{4}}\right)\right].\end{equation*}
\end{corollary}

%
\subsection{Epstein type moments}

We now consider the sequence $\gL = \{ n_1^2+n_2^2+\ldots+n_d^2|\vec
n \in \B{N}_0^d-\{\vec 0\}\}$. Using the resummation formula
\cite{hille} $$\sum_{l=-\infty}^\infty e^{-t l^2} = \sqrt{\frac \pi t}
\sum_{l=-\infty} ^\infty e^{-\frac{\pi^2} t l^2}, $$ it is easily
seen that the associated partition function has the small-$t$
asymptotic expansion
\begin{equation*} \gQ_\C{E}(t)\sim\sum_{n=0}^d A_{-\frac n 2}
t^{-\frac{n}{2}}\end{equation*}
where the $A_{-\frac n 2}$ are given by $$A_0 = \frac 1 {2^d} -1,
\quad \quad A_{-\frac n 2} = \frac 1 {2^d} {d \choose n} \pi
^{\frac n 2}, \quad n=1,\ldots,d.$$
With the help of (\ref{Gen Res}) this determines the relevant
residues. Additional quantities needed in the corollaries stated
below  can be extracted from
\cite{elin,eliz94b,epst03-56-615,epst07-63-205,KirK1}.

We now approach two and three-dimensional Epstein type moments in
the same way that we considered the Barnes type.

\subsubsection{Two-dimensional}

Note that for the two-dimensional case, \eqref{Gen n(a)} gives
\begin{equation*} \ga=
\left(\frac{\pi\gz_\C{R}(2)}{4}\right)^{\frac{1}{2}}n^{-\frac{1}{2}}+\left(\frac{\pi^{\frac{1}{4}}
\gz_\C{R}\left(\frac{3}{2}\right)}{4\sqrt{2}(\gz_\C{R}(2))^{\frac{1}{4}}}\right)
n^{-\frac{3}{4}}-\frac{3}{8n}+\C{O}\left(n^{-\frac{5}{4}}\right).
\end{equation*}

Using this information we obtain the following corollaries to the
General Moment Theorems.

\begin{corollary} For the two-dimensional Epstein type sequence with
$c=0$ and $\vec{r}=\vec{1}$,
\begin{multline*}
t_{\C{E}_2}^0(n)=\left(\pi^2\gz_\C{R}(2)\right)^{\frac{-1}{2}}
\left(\frac{\pi\gz_\C{R}(2)}{4n}\right)^{\frac{9}{8}}\\
\times\exp\left[\sqrt{\pi\gz_\C{R}(2)n}+\left(\frac{\pi^{\frac{1}{4}}\gz_\C{R}\left(\frac{3}{2}\right)}
{4\sqrt{2}(\gz_\C{R}(2))^{\frac{1}{4}}}\right)n^{\frac{1}{4}}+\frac{\left(\gz_\C{R}\left(\frac{3}{2}\right)\right)^2}{
32\gz_\C{R}(2)}+\gz_{\C{E}_2}'(0)\right]\\
\times\left[1+\C{O}\left(n^{-\frac{1}{4}}\right)\right].
\end{multline*}
\end{corollary}

\begin{corollary} For the two-dimensional Epstein type sequence with
$c=0$, $\vec{r}=\vec{1}$, and $k\geq 1$,
\begin{multline*}
t_{\C{E}_2}^k(n)=\left(\pi^2\gz_\C{R}(2)\right)^{-\frac{1}{2}}
\left(\frac{\pi\gz_\C{R}(2)}{4n}\right)^{\frac{9-4k}{8}}\\
\times\exp\left[\sqrt{\pi\gz_\C{R}(2)n}+\left(\frac{\pi^{\frac{1}{4}}\gz_\C{R}\left(\frac{3}{2}\right)}
{4\sqrt{2}(\gz_\C{R}(2))^{\frac{1}{4}}}\right)n^{\frac{1}{4}}+\frac{\left(\gz_\C{R}\left(\frac{3}{2}\right)\right)^2}{
32\gz_\C{R}(2)}+\gz_{\C{E}_2}'(0)\right]\\
\times\sum\left(\fp{t}{1}{\gz_{\C{E}_2}(t)}-\frac{\pi}{8}
\log\left(\frac{\pi\gz_\C{R}(2)}{4n}\right),\gz_{\C{E}_2}(2),2\gz_{\C{E}_2}(3),
\ldots,(k-1)!\gz_{\C{E}_2}(k)\right)\\ \times
\left[1+\C{O}\left(n^{-\frac{1}{4}}\right)\right].
\end{multline*}
\end{corollary}

Theorem \ref{Gen m} gives the expected number of summands.

\begin{corollary} For the two-dimensional Epstein type sequence with
$c=0$ and $\vec{r}=\vec{1}$,
\begin{equation*}
m_{\C{E}_2}(n)=\left(\frac{4n}{\pi\gz_\C{R}(2)}\right)^{\frac{1}{2}}\left(
\fp{t}{1}{\gz_{\C{E}_2}(t)}-\frac{\pi}{8}\log\left(\frac{\pi\gz_\C{R}(2)}{4n}\right)
\right)\left[1+\C{O}\left(n^{-\frac{1}{4}}\right)\right].\end{equation*}
\end{corollary}

%

\subsubsection{Three-dimensional}

For the three-dimensional Epstein case, \eqref{Gen n(a)} gives
\begin{multline*} \ga=
\left(\frac{3\pi^{\frac{3}{2}}\gz_\C{R}\left(\frac{5}{2}\right)}{16}\right)^{\frac{2}{5}}n^{-\frac{2}{5}}+
\left(\frac{3^{\frac{3}{5}}\gz_\C{R}(2)}{5}\right)\left(\frac{\pi}{2\gz_\C{R}\left(\frac{5}{2}\right)}\right)^{\frac{2}{5}}
n^{-\frac{3}{5}}\\
+\frac{3^{\frac{4}{5}}\pi^{\frac{1}{5}}}{2^{\frac{1}{5}}100\left(\gz_\C{R}\left(\frac{5}{2}\right)\right)^{\frac{6}{5}}}
\left(5\gz_\C{R}\left(\frac{3}{2}\right)\gz_\C{R}\left(\frac{5}{2}\right)
-2(\gz_\C{R}(2))^2\right) n^{-\frac{4}{5}}-\frac 7 {20n}
\C{O}\left(n^{-\frac 6 5}\right).
\end{multline*}

Here we may note that $\mu_0=\frac{3}{2}>1$, so that we have the
following corollary to the General Moment Theorems.

\begin{corollary} For the three-dimensional Epstein type sequence with
$c=0$, $\vec{r}=\vec{1}$, and $k\in\B{N}_0$,
\begin{multline*} t_{\C{E}_3}^k(n)
=\left(\frac{15}{16}\pi^{\frac{5}{2}}\gz_\C{R}\left(\frac{5}{2}\right)\right)^{-\frac{1}{2}}
\left(\frac{3\pi^{\frac{3}{2}}\gz_\C{R}\left(\frac{5}{2}\right)}{16n}\right)^{\frac{27-12k}{20}}
\left(\frac{\pi^{\frac{3}{2}}}{8}\gz_\C{R}\left(\frac{3}{2}\right)\right)^k\\
\times\exp\left[\left(\frac{5\pi^{\frac{3}{5}}\left(\gz_\C{R}\left(\frac{5}{2}\right)\right)^{\frac{2}{5}}}
{6^{\frac{3}{5}}2}\right)n^{\frac{3}{5}}+\frac{3^{\frac{3}{5}}\gz_\C{R}(2)}{2}\left(
\frac{\pi}{2\gz_\C{R}\left(\frac{5}{2}\right)}\right)^{\frac{2}{5}}n^{\frac{2}{5}}\right.\\
\left. +\frac{
3^{\frac{4}{5}}\pi^{\frac{1}{5}}}{2^{\frac{1}{5}}20\left(\gz_\C{R}\left(\frac{5}{2}\right)\right)^{\frac{6}{5}}}
\left(5\gz_\C{R}\left(\frac{3}{2}\right)\gz_\C{R}\left(\frac{5}{2}\right)-2(\gz_\C{R}(2))^2\right)n^{\frac{1}{5}}\right.\\
\left.
+\frac{(\gz_\C{R}(2))^3}{10\left(\gz_\C{R}\left(\frac{5}{2}\right)\right)^2}-
\frac{3\gz_\C{R}\left(\frac{3}{2}\right)\gz_\C{R}(2)}{20\gz_\C{R}\left(\frac{5}{2}\right)}+\gz_{\C{E}_3}'(0)\right]
\left[1+\C{O}\left(n^{-\frac{1}{5}}\right)\right].
\end{multline*}
\end{corollary}

Theorem \ref{Gen m} yields the following corollary.

\begin{corollary} For the three-dimensional Epstein type sequence with
$c=0$ and $\vec{r}=\vec{1}$,
\begin{equation*}
m_{\C{E}_3}(n)=\frac{\pi^{\frac{3}{2}}\gz_\C{R}\left(\frac{3}{2}\right)}{8}
\left(\frac{16n}{3\pi^{\frac{3}{2}}\gz_\C{R}\left(\frac{5}{2}\right)}
\right)^{\frac{3}{5}}
\left[1+\C{O}\left(n^{-\frac{1}{5}}\right)\right].\end{equation*}
\end{corollary}

%

\section{Conclusion}

In this article we have developed a systematic approach to obtain,
for large $n$, the moments of the number of partitions of $n$ into
$m$ summands from a given sequence $\gL$ of natural numbers. Under
suitable assumptions on $\gL$, see \eqref{Gen Asym}, the moments
are expressed in terms of information obtained from the small-$t$
asymptotic expansion of the associated partition function. The
main technical tool is the saddle-point method in combination with
the application of \eqref{(e^-z) int}.

Whenever the small-$t$ asymptotic expansion of the partition
function can be easily determined, the large-$n$ expansions of the
moments can be easily found. This has been demonstrated in Section
\ref{applications} with sequences given by linear and quadratic sums of integers.

\section*{\bf Acknowledgments} KK acknowledges support by the
NSF through grant PHY-0757791.


\end{document}